\documentclass{article}

\usepackage[square,sort,comma,numbers]{natbib}

\usepackage{amsmath}
\usepackage{graphicx}
\usepackage{arxiv}

\usepackage[utf8]{inputenc} 
\usepackage[T1]{fontenc}    
\usepackage{hyperref}       
\usepackage{url}            
\usepackage{booktabs}       
\usepackage{amsfonts}       
\usepackage{nicefrac}       
\usepackage{microtype}      
\usepackage{lipsum}

\title{A Short Survey of Topological Data Analysis in Time Series and Systems Analysis}
\author{
  Shafie Gholizadeh \\
  Department of Computer Science \\
  University of North Carolina at Charlotte\\
  Charlotte, NC 28223 \\
  \texttt{sgholiza@uncc.edu} \\
   \And
 Wlodek Zadrozny \\
  Department of Computer Science\\
  University of North Carolina at Charlotte\\
  Charlotte, NC 28223 \\
  \texttt{wzadrozn@uncc.edu} \\
}

\begin{document}
\maketitle

\begin{abstract}
Topological Data Analysis (TDA) is the collection of mathematical tools that capture the structure of shapes in data. Despite computational topology and computational geometry, the utilization of TDA in time series and signal processing is relatively new. In some recent contributions, TDA has been utilized as an alternative to the conventional signal processing methods. Specifically, TDA is been considered to deal with noisy signals and time series. In these applications, TDA is used to find the shapes in data as the main properties, while the other properties are assumed much less informative. In this paper, we will review recent developments and contributions where topological data analysis especially persistent homology has been applied to time series analysis, dynamical systems and signal processing. We will cover problem statements such as stability determination, risk analysis, systems behaviour, and predicting critical transitions in financial markets.
\end{abstract}


\section{Introduction}\label{sec:introduction}
The main target in Topological Data Analysis (TDA) is to find the shape or the underlying structure of shapes in data. In its recent contribution, TDA is often considered to deal with a huge number of discrete points in high-dimensional space., i.e., a data cloud. A simple way to show that how a large number of data points can represent the whole space is to construct the Voronoi Diagram \cite{aurenhammer1991voronoi}. Voronoi diagram partitions the space to convex sub-regions (e.g., polygons, polyhedrons, etc) and each sub-region covers the area in which a site (i.e., a particular data point) is the closest data point. TDA methods usually introduce a similar intuition, but follow more complex methodologies. While the geometry (as offered by Voronoi Diagrams) depends on distances to study shapes, topology benefits from homeomorphisms that are closed with respect to stretching or shrinking. Therefore, TDA methods are much less sensitive to the choice of metric  \cite{zomorodian2012topological}. This will allow us to use TDA for a wide family of problems in which the shape of data matters. We will follow some recent developments in TDA and will focus on a widely used topological tool called Persistent Homology.

As we discussed, some large data sets may contain a huge number of discrete points. Here, TDA offers a tool to interpret these records as a meaningful shapes. A precise yet easy way to describe this technique is to compare it with human visual interpretation, like looking at the close stars in the sky and distinguishing a continuous shape of a bear \cite{gholizadeh2018topological}. In TDA, this procedure is often automated by a technique called persistent homology \cite{carlsson2009topology, edelsbrunner2008persistent, ghrist2008barcodes, chen2015mathematical}. Persistent homology deals with the number of holes in the shape at each dimension, when we try to connect close discrete points. 

Recently, TDA has been considered as an alternative to the traditional data analysis and machine learning algorithms. Recent contribution of TDA covers a wide range of applications. Pachauri et al. in \cite{pachauri2011topology} utilized TDA on a classification problem of Alzheimer detection. Guss and Salakhutdinov in \cite{guss2018characterizing} showed how topological data analytic can be hired to select task-optimal architecture of neural networks. De Silva and Ghrist in \cite{de2006coordinate} and \cite{de2007coverage} used homology in analysis of sensor network coverage where the sensors have some week capabilities. The idea then developed in many other works \cite{adams2015evasion,carlsson2010zigzag,ghrist2017positive}. Stolz et al. in \cite{stolz2016topological} studied the topological shape of Brexit referendum in United Kingdom. In this paper, we will focus only on a specific application of TDA, where persistence homology is used to analyze the time-variant characteristics of data. In section \ref{theo}, we will review basic definitions and required tools from topology that are directly or indirectly utilized in our selected applications surveyed in section \ref{app}. While the origin of the TDA lies in mathematics, even our focused application covers different scientific fields like economy, finance, control and systems engineering, and statistics. We will summarize these applications in two categories. In subsection \ref{systems}, we will look at the applications of TDA in signal processing as well as systems and control engineering. Subsection \ref{finance} reviews the contribution of TDA in finance and financial econometrics.


\section{Topology and Data}
\label{theo}
A topology on a set is defined by a collection of subsets containing the empty set and the whole set that is closed under intersection and union operations. By definition, topology is not generally closed under complementing. Instead, we call any element in a topology and the complement of any element in topology an \emph{Open Set} and a \emph{Closed Set} respectively.

\begin{figure*}[!ht]
\centering
\includegraphics[width = 0.8\linewidth]{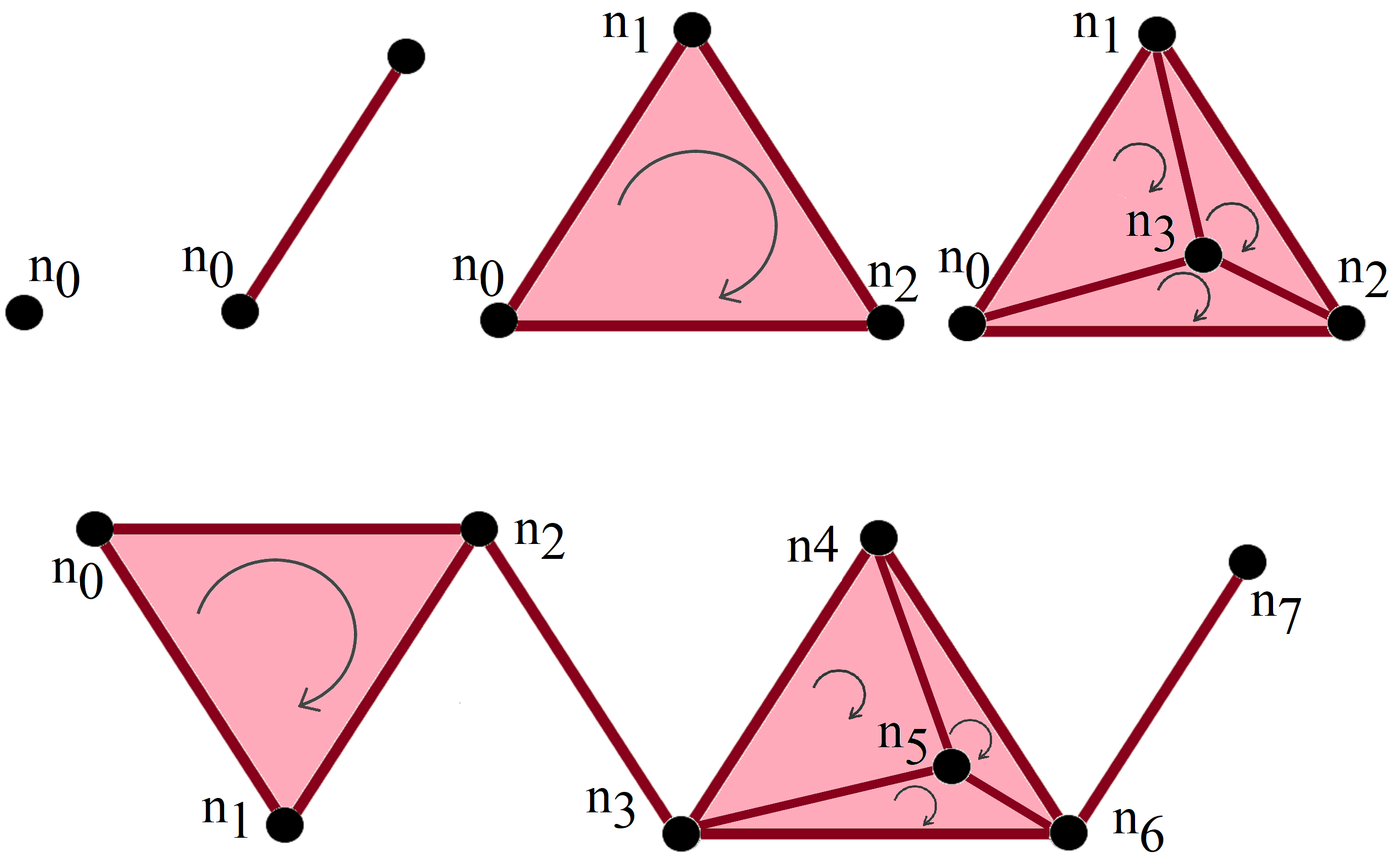}
\caption{$k$-simplices for $k =$ $0$, $1$, $2$, and $3$ (top). A simplicial complex (bottom).}
\label{fig:Kcomplex}
\end{figure*}

In TDA, we often use \emph{Simplicial Complexes} to study the shapes. A \emph{Simplex} can be a single point (0-simplex), two connected data points (1-simplex), tree fully-connected points (2-simplex), or generally $k+1$ fully-connected data points ($k$-simplex). For consistency of the definition, we use $(-1)$-simplex to describe an empty set \cite{zomorodian2010computational}. Then we can define a simplicial complex as the union of some simplices if it has one conditions: If a simplex is in the simplicial complex, any of its subsets should be also be in the complex. Some examples of $k$-simplices and a simplicial complex are illustrated in Figure \ref{fig:Kcomplex}. 

On a high dimensional space where the the dimension of simplices could be high, we usually need to reduce the dimensions. Moving from the highest dimensions and decreasing the dimensions step-wise, we will get a Chain Complex ($C_n:  n=k, k-1, \dots , 1$) where $C_K$ is the set of all $k$-simplices. Then we can define the \emph{Euler Characteristic} for a data cloud (here the simplicial complexes), as in Equation \ref{eq:EQ1} \cite{bass1976euler, zomorodian2012topological}. 

\begin{equation}
\begin{split}
    \chi(C)&=\sum_k\ (-1)^k \  dim \  C_k \\
           &=\sum_k\ (-1)^k \ \text{number of k-simplices}\\
           &=\sum_k\ (-1)^k \ dim \ H_k 
\label{eq:EQ1}
\end{split}
\end{equation}

We define the $k^{th}$ chain group $C_k$ as the free Abelian group of oriented $k$-simplices. Thus, any element $c$ in $C_k$ is a $k$-chain satisfying $c=\sum_i c_i[\sigma_i]$, where $\sigma_i$ is any $k$-simplex and $c_i \in \mathbb{Z}$ is a coefficient. Now, we can define the \emph{Boundary Homomorphism}  $h_k: C_k \rightarrow C_{k-1}$ as a homomorphism defined on any simplex in $C$ \cite{zomorodian2010computational,zomorodian2012topological}. Note that homomorphism is a continuous one-by-one map between two topological space that has also a continuous inverse.

 \begin{equation}
  h_k[v_0 , \dots , v_k] =\ \sum_i (-1)^i \ [v_0 , \dots , \hat{v_i} , \dots , v_k]
  \label{eq:EQ2}
 \end{equation} 
 
In Equation \ref{eq:EQ2}, $\hat{v_i}$ denotes the vertex that is deleted from the sequence of vertices. $H_k$ in Equation \ref{eq:EQ1} is the $k^{th}$ \emph{Homology Group} defined as:
\begin{equation}
    H_k= {Kernel(h_k)\ / \ Image(h_{k+1})}
\label{eq:EQ3}
\end{equation}

In Equation \ref{eq:EQ1}, $dim C_k$ is equal to the $k^{th}$ \emph{Betti Number}. The $i^{th}$ Betti number is defined as the number of $i$-dimensional holes a in simplicial complex. More specifically, $\beta_0$ is the number of connected component, $\beta_1$ is number of $1$-$D$ holes and $\beta_2$ is the number of $2$-$D$ voids, etc. Note that For a $m$-dimensional shape, for any $n \geq m$, $n^{th}$ Betti number is zero. Betti numbers for some topological shapes are shown in Table \ref{tab:betti}.

\begin{table*}[ht]
    \caption{Betti numbers for some basic shapes.}
\label{tab:betti}
    \centering
    \resizebox{0.55\linewidth}{!}{
    \begin{tabular}{|c|c c c c|}
    \hline
    Betti Number & A Point & Circle & Sphere & Torus \\ \hline
    $\beta_0$ & 1 & 1 & 1 & 1 \\
    $\beta_1$ & 0 & 1 & 0 & 2 \\
    $\beta_2$ & 0 & 0 & 1 & 1 \\
    $\beta_3$ & 0 & 0 & 0 & 0 \\
    $\vdots$ & $\vdots$ & $\vdots$ & $\vdots$ & $\vdots$ \\\hline
    
    \end{tabular}
    }
\end{table*}

\begin{figure*}[!ht]
\centering
\includegraphics[width=0.99\linewidth]{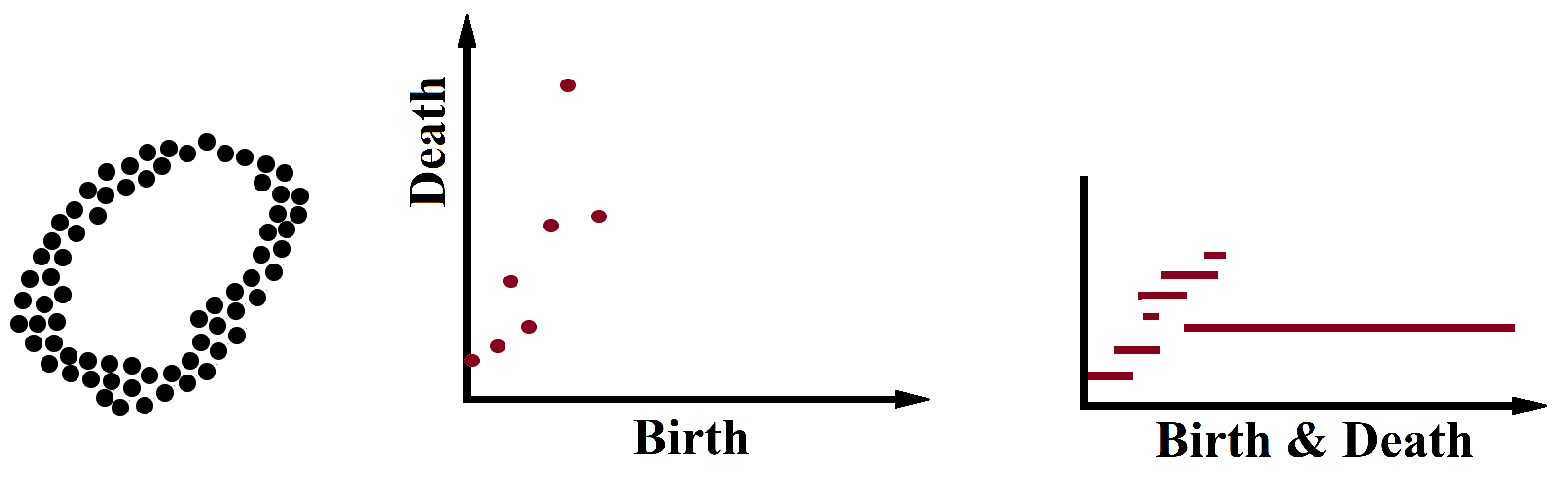}
\caption{Some data points in $2$-$D$ space (left), the persistence diagram (middle), and its equivalent barcode (right).}
\label{fig:Barcode}
\end{figure*}

\emph{Persistent Homology} is a technique in TDA to find topological patterns of the data \cite{edelsbrunner2000topological,zomorodian2005computing,munch2017user}. Dealing with a set of discrete data points, we can define a radius around each data point and connect each pair of points within that radius of each other. Then we may observe many holes or loops in the resulted simplicial complex. However some data points might construct a fully-connected partition where there is no hole, i.e., a $k$-simplex. If we increase the defined radius gradually, of in fact decrease the resolution, obviously the resulted simplicial complex will change. Subsequently, the holes and their numbers (Betti numbers) in the shape will change. So, increasing the radius, many holes (e.g., loops in dimension $1$) will come to the picture and then will disappear. We may illustrate the birth and the death radii of the hoes for each dimension in \emph{Persistence Diagram} . Equivalently, the birth  and the death radii of holes might be shown  with barcodes where the lifetime of every hole is represented by a one dimensional bar from the birth radius to the death radius \cite{collins2004barcode,ghrist2008barcodes,carlsson2014topological}. An example of these barcodes and the equivalent persistence diagram are shown in Figure \ref{fig:Barcode}.

In persistent homology, we can study the distances among data points in different ways. In the procedure we described above, the information structure based on thresholding distances is called \emph{Vietoris-Rips Filtration} \cite{ghrist2008barcodes}. In a Vietoris-Rips complex, any $k$-simplex is consisted of $k$ nodes whose pairwise distance is less than or equal to the threshold. One may consider a slightly stronger constraint in which the regions around the nodes of a simplex within the radii equal to the threshold altogether should have a non-empty intersection. In this case the result is called \emph{\v{C}ech Complex}. Dealing with a weighted graph (instead of a data cloud), we may threshold the weights and increase the weights threshold gradually. In this case, the structure is called \emph{Weight Rank Clique Filtration} \cite{petri2013topological}.

\begin{figure}[!ht]
\centering
\includegraphics[width=0.99\linewidth]{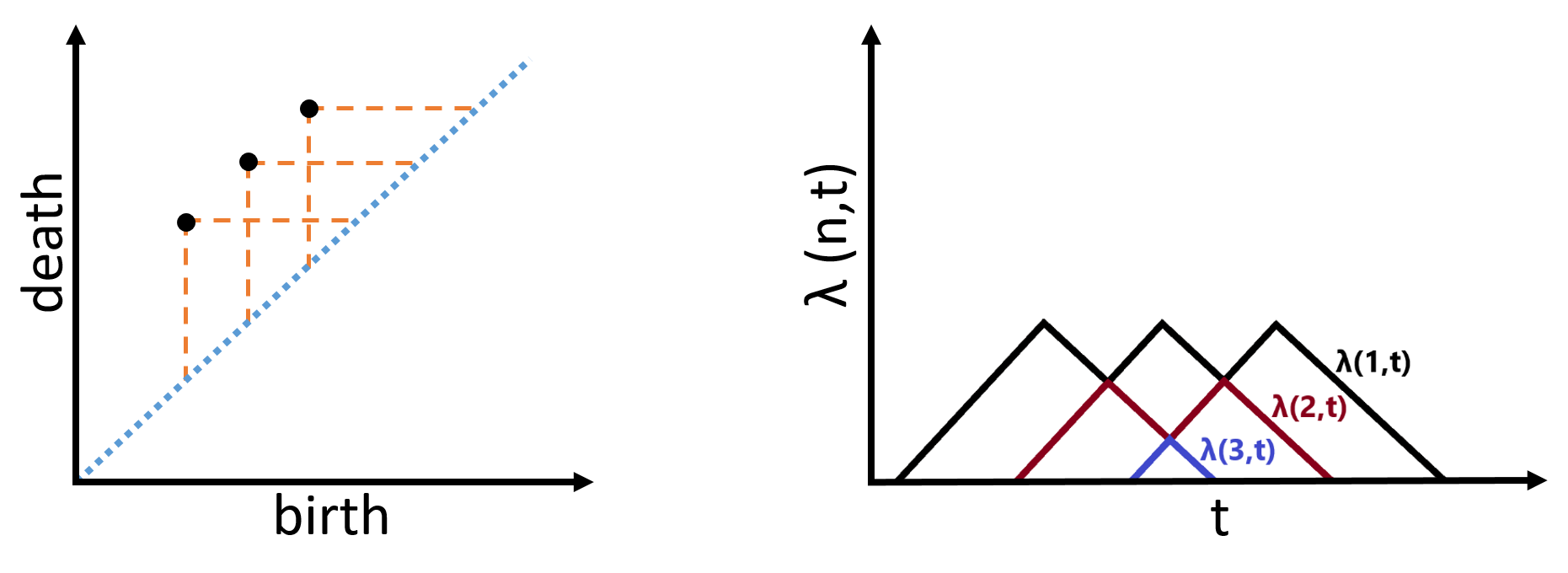}
\caption{A persistence diagram (left) and its equivalent persistent landscape (right).}
\label{fig:persistentLandscape}
\end{figure}

Bubenik in \cite{bubenik2015statistical} showed that a the topological signatures in data can be also summarized in a real-valued function called \emph{Persistent Landscape}. Let $\beta(birth,death)$ denote the Betti number of a module at some particular dimension $d$ ($d^{th}$ Betti number) in the resolution interval ($birth$ , $death$). Then the persistent landscape is a real-valued function $\lambda$: $\mathbb{N} \times \mathbb{R} \rightarrow \mathbb{R}$ such that $\lambda(n,t) = sup\{radius \geq 0 \text{ } | \text{ } \beta(t-radius \text{ } , \text{ } t+radius) \geq n\}$ for any $n \in \mathbb{N}$. Intuitively, if we connect each of the points in the persistence diagram to the bisector line of $\mathbb{R}^2$ via horizontal and vertical lines, and rotate the space by $\pi/4$ clockwise, the result is the landscape function. Persistent Landscape for an equivalent persistence diagram is shown in Figure \ref{fig:persistentLandscape}. Note that we may calculate or even estimate persistent landscape for different samples of data cloud where the mean of results on different samples still captures the topological signature.

\section{The Applications}
\label{app}
Here we will review recent developments and contributions where topological data analysis especially persistent homology has been applied to time series analysis, dynamical systems and signal processing. More precisely, the list of problem statements includes stability determination, systems behaviour, prediction of bifurcations, structural breaks and critical transitions, and catastrophe risk analysis in stock markets.

\subsection{TDA in Systems Engineering and Signal Processing}
\label{systems}

Skraba et al. in \cite{skraba2012topological} showed that persistent homology can be used in time delay embedding models. Time delay embeddings translate a $1$-dimensional time series to a $d$-dimensional time series in which the current value at each time with ($d-1$) lags coordinate \cite{packard1980geometry, takens1981detecting}. Skraba et al. developed a framework of analyzing dynamic systems based on topological data analysis that requires almost no prior information of the underlying structure. Instead, a discrete sample of data point is being studied in periodic, quasi-periodic and recurrent systems. The same approach have been considered frequently in the later works on different applications of time series. The authors suggested that clustering $d$-dimensional delay embedding of a time series in some subspaces of $\mathbb{R}^d$ can easily reveal recurrent nature of the system on appeared loops and returning paths. The next step is to utilize persistent homology or more precisely the  persistence diagram or Betti numbers to measure these loops. Note that these loops in delay-coordinate embedding do not necessarily exist in the samples of the initial time-series. This is one of the reasons that the authors preferred to work on delay embedding $\mathbb{R}^d$ field. In addition, tuning the time delay parameter will result in a more robust model. Based on Vietoris-Rips or any similar filtration, one can construct the persistence diagram of the delay embedding. As the results, a periodic system will end up with the Betti numbers (or equivalently persistence diagram) of a circle. Similarly, a quasi-periodic system with $n$ periods will end up with the persistent Betti number of an $n$-dimensional torus. For example when $n=2$ the system will have the Betti numbers of the torus in Table \ref{tab:betti}. Finally, the persistence diagram of a recurrent system will look like that of a bouquet of circles. Note that in the first two cases, after extraction of persistence diagrams there exist many easy ways to retrieve the time periods explicitly. The road map is simple. Each time period is reflected in the sizes of loops or holes in the space of delay embedding, where persistence homology has an easy job to measure them. Thus, in addition to the system type identification or counting the loops (equivalently the number of periods), this approach is capable of measuring the specification of periods.

Perea and Harer in \cite{perea2015sliding} also worked on the same problem, using the topology of sliding window to measure the strength of persistent homology in periodic signals. The authors showed more precisely how persistent homology can be used to detect periodic behavior of signals. They used $1$-dimensional persistence diagrams of sliding window on periodic noisy signals to discover periodicity and showed how their approach can comparably perform as the state of the art. Perea in \cite{perea2016persistent} extended the theories in \cite{perea2015sliding} to the quasi-periodic functions (i.e., linear summation of periodic functions with irrelevant frequencies). The author proved the way to obtain the optimal choice of time delay and window size for for sliding window embedding of such functions and then calculated the upperbound and lowerbound for persistent homology of sliding window.

Berwald and Gidea in \cite{berwald2013critical} used persistence diagrams to detect critical transitions in genetic regulatory systems which have stochastic characteristics. Based on Vietoris-Rips filtration, the authors built persistence diagram of time windows in data and then tried to reveal qualitative changes based on significant topological difference in consecutive windows. They suggested whenever a critical transition is going to happen, the persistence diagrams of time windows illustrate different properties in the term of points distribution. The idea was expanded in \cite{berwald2013automatic} where Berwald et al. suggested a novel approach to distinguish different regimes in a time-dependent dynamical system. Constricting the barcodes in dimension $1$, the authors used $k$-means algorithm to cluster the time windows with the most significant bars. Intuitively, if a bar is significantly longer that all the other bars, one may assume that the barcode suggests a periodic regime. There also exist similar conditions to recognize quasi-periodic and recurrent regimes as discussed in \cite{skraba2012topological}. But here the authors feed the barcodes of different time windows to an unsupervised algorithm (e.g., $k$-means). As the result, local bifurcations are recognizable where the window size is relatively small, or when the cluster tag is changing frequently over the time. On the other hand, global bifurcation is detectable when the cluster tag is changing only once or more precisely when the distributions of cluster tags are significantly different before and after a certain time.

Garland et al. in \cite{garland2016exploring} showed how the \emph{Witness Complex} may be used to explore the underlying topology of noisy sample in dynamical systems. Here the intuition is similar to \cite{berwald2013automatic}, except the fact that to build the persistence diagrams, the authors utilized \emph{Witness Complexes} \cite{de2004topological} instead of Vietoris-Rips complexes. Witness Complexes are often in smaller size and therefore easier to handle. In a weak Witness complex a subset (called \emph{landmark}) $LM$ of data cloud $DC$ is chosen. A simplex $Sim=\{l_1,\dots,l_n\}$ has a \emph{weak witness} $\omega \in P$ if and only if for any $(l^*,r) \in Sim \times (LM\backslash{Sim})$ we have $Distance(l^*,\omega) \leq Distance(r,\omega)$. While choosing landmark subset via some heuristic reduces the computational cost by far, witness complexes will still cover useful topological features. Intuitively, witness complex is the output of pruning a larger complex, while the principal properties are carefully relayed through the pruning steps. Mittal and Gupta in \cite{mittal2017topological} followed this idea and tried to use persistence diagrams from witness complexes to describe and detect chaos and bifurcations in time series. The authors introduced six different useful features from the persistence diagram including number of holes, average lifetime of holes, maximum diameter of holes and maximum distance between holes in each dimension. Useful feature chosen upon the feature selection training may feed the descriptive or predictive models. Note that the idea of feeding topological features of systems to a learning model was introduced in \cite{skraba2012topological} where the barcodes were used in $k$-means algorithm. Then the distribution of cluster tags was analyzed over the time. But here in \cite{mittal2017topological}, different features are extracted from the persistence diagrams and then the learning model decides whether/how to use those features. The other capability is to feed the features alone or along with other non-topological features to the models (e.g., regression, classification, and clustering algorithms).

In \cite{chazal2013persistence}, Chazal et al. developed the idea of persistence-based clustering. In mode-seeking clustering, the local peaks of density function are discovered via some hill-climbing method and being used as the cluster centers. Since these centers are often unstable, the authors suggested that stable peaks can be found in the terms of persistence. While any local peak will present a unique a point to the persistence diagram, insignificant peaks (e.g., peaks that appeared due to sampling limitations) are to be eliminated. This is a trivial problem on persistence diagram. Only the most important peaks can have a long lifetime (i.e., the difference between the birth and the death radius in Vietoris-Rips filtration). So, the lifetime of points in the persistence diagram reveals an underlying hierarchy for centroids importance. Morevere the persistence diagram can suggest the number of significant points and the efficient number of clusters. Chang et al. in \cite{chang2013persistent} proposed Multi-Persistent Clustering analysis to cluster molecular dynamics simulation data based on scale and density. Similar to \cite{chazal2013persistence}, the clusters are defined in the terms of persistence, but here the authors measured persistence on a $2D$-space based on scale and density of data points. Pereira and de Mello in \cite{pereira2015persistent} also followed the general idea in \cite{chazal2013persistence} but more focused on topological properties each cluster and the relations among them. Their proposed method uses two dimensional delay embedding of time series. The authors showed that how different types of time and the resulted embedding  will result in different barcodes.

Khasawne and Munch in \cite{khasawneh2014stability} utilized the idea of measuring persistent homology on time delay embedding to evaluate the stability of time-dependent systems in which the main characteristic is defined by Stochastic Delay Differential Equations. When the delay and the stochastic terms simultaneously exist in the system, it is usually difficult to summarize the behavior and determine the stability, though in special cases there are some methods that combine stochastic calculus and numerical methods to evaluate SDDEs (e.g., in It{\^o} sense). Here the authors used a delay embedding of the signal similar to \cite{skraba2012topological} to construct persistence diagrams and distinguish the system conditions (i.e., equation parameters) under which the system is stable. They showed that even for the systems with delay and stochastic characteristics, persistence homology is useful for detecting bifurcations and qualitative behavioral change. 

Emrani et al. in \cite{emrani2014persistent} used persistent barcodes in dimension $1$ (equivalent to the first Betti number) for an efficient algorithm to distinguish wheezes among signals. Assuming the frequencies of the wheeze signals to be piecewise constant functions of time, a wheeze signal can be represented as the summation of some sinusoidal functions of time. It can be theoretically illustrated that for the delay embedding of such summation, there is at least one persistent barcode in dimension $1$. The authors used this simple fact and proved that for $2$-dimensional delay embedding of a wheeze time-series, the first persistent Betti number is at least one.  On the other hand, their experimental results suggest that the first persistent Betti number for delay embedding of non-wheeze signal is zero. In fact there is a huge difference between the barcodes of wheeze and non-wheeze time-series as shown in their work. The authors also showed that a relatively small subsample of the signal is enough to distinguish wheezes. This makes their algorithm efficient efficient enough to run on large data sets in the real world applications.

Sanderson et al. in \cite{sanderson2017computational} suggested that persistent homology may capture the difference when the same musical note is played on some different instruments. The authors used persistence diagrams of $2$-dimensional delay embedding to distinguish musical instrument. They trained a classifier on a few labeled diagrams of different instruments. The classifier performed much better than a traditional classifier based on Fast Fourier Transform of time series. Instead of feeding the persistence diagram directly to the classifier, the authors used Persistent Rank Functions-- i.e., functions that capture the number and position of points on persistence diagrams \cite{robins2016principal}. 

Gholizadeh et al. in \cite{gholizadeh2018topological} define a framework to process textural documents as $n$-dimensional time series of $n$ different entities. After extracting a list of important entities, the authors used the set of indices in the text where each entity is appearing. Then persistent homology is used to interpret the co-appearance of different entities. For each pair of entities, a difference is defined by Wasserstein distance of the two indices vectors. Based on such distances, a context-free graph of entities is constructed for each document. In any of these graphs, vertices represent entities and the links represent their distances. The authors showed that persistent homology is capable of extracting topological signature of the author from these graphs. They used persistent diagram of these graphs to predict the author in binary classifications where the average accuracy was $77\%$. As for classification, the author once again used Wasserstein distance to measure the difference among persistence diagrams of different documents. Having all distances among persistence diagrams, a distance-based algorithm was used to predict the authors.


\subsection{TDA in Financial Econometrics}
\label{finance}

Predicting the structural breaks and catastrophe stock market crash as early as possible is considered one of the most valuable yet difficult tasks in financial domain. Gidea in \cite{gidea2017topological} utilized the ideas of \cite{perea2015sliding,perea2016persistent} to detect early signs of critical transitions in financial time series using persistence homology. in financial market, the time series of each stock of an index may depend on some particular condition but on the other hands these indices often respond to mutual information, announcements and regulatory entities. That is why the author assumed that the topology of their correlation graph matters. Considering the time series of multiple stock returns at each instant time as the nodes of a weighted graph, cross correlations among the nodes are being assumed as the weights of the graph. So at each time, there exist a weighted graph representing the correlations of stock returns. To get the persistence diagram for each of the graphs, the author used weight rank clique filtration which thresholds the weights and increase the weights stepwise. Finally, the differences between each persistence diagram and the persistence diagram at the initial time will construct the time series of differences. Here the differences are calculated based on \emph{Wasserstein distance}\cite{edelsbrunner2010computational} which measures the minimum cost to map a distribution to another one. Assuming that the persistence diagram is robust in normal condition, the time-series of distances can represent and reveal the loss of normal market conditions. The author applied the proposed method to the data set of companies in Dow Jones Industrial Index at time interval including 2007-2008 financial crises and the early signs appeared eight months before the last stock market peak in October 2008. Here the approach is much similar to what was previously done in \cite{nobi2014correlation} except the idea of using persistent homology. Nobi et al. in \cite{nobi2014correlation} worked on time series of global and local stock indices, built the correlation graph from of these time series in sliding window and got into a sequence of threshold graphs. However here the authors tried to measure the topological change in the sequence of correlation graphs based on Jaccard index between any two consecutive graphs and focused on degree distribution of the nodes in the graphs that reveals important information regarding network density.

Gidea and Katz in \cite{gidea2018topological} utilized persistent landscape to predict crashes on financial time series. Their method is similar to the general approach in \cite{gidea2017topological}. Here the authors used $L^p$-$norms$ (i.e., Minkowski distance of order p from the origin of a Euclidean space) of the persistence landscape to detect the signs of a crash. A financial crash often happens after a period of high variance in is market indices and high cross-correlations among stocks. Thus, intuitively the $L^p$-norms may have a rising trend prior to the crash. The authors used S\&P 500, DJIA, NASDAQ, and Russell 2000 data and analyzed their daily return prior to dotcom crash on 2000 and also 2008 financial crisis. Four indices make a $4$-dimensional time series that was analyzed in sliding window of a fixed size with sliding step set to one day. Calculating the $L^p$-norm for each dimension $n$ of the persistent landscape $\lambda(n,x)$ within each window, we will have a vector of $L$-norms. Finally the $L^p$-norm of this vector is a function of time that is used for analysis purpose. As the results, $L^1$ and $L^2$ norms illustrated significant rising trends before the crashes.


\subsection{A Brief Sketch of the Other Related Works}
\label{other}

Topological data analysis has been applied to a few other problem statements in time series and dynamical systems analysis. Maleti{\'c} et al. in \cite{maletic2016persistent} studied the persistence diagrams of delay embedding in dynamical system similar to \cite{skraba2012topological} and \cite{berwald2013automatic}, and categorized the expected results on many well-known non-linear systems of chaotic behavior. Topaz et al. in \cite{topaz2015topological} used persistent homology to study biological aggregations (i.e., dynamical systems defined by a scholar time-series for multiple agents who interact and influence each other. Venkataraman et al. in \cite{venkataraman2016persistent} used persistence diagrams of time delay embedding for human action recognition based on three dimensional motion capture data. Labeling a few diagrams, they showed how persistent homology can outperform the state of the art in classifying action-- e.g., jump, run, sit, dance, or walk. Perea et al. in \cite{perea2015sw1pers} used persistent homology of sliding window for periodicity detection in time series of gene expression. Seversky et al. in \cite{seversky2016time} provided a collection of persistence diagrams from a variety of well-known time series sources that can be used for future studies (e.g., feeding the persistence diagrams to any unsupervised and even sometimes supervised algorithm in traditional machine learning). Stolz et al. in \cite{stolz2017persistent} constructed wight rank clique filtration from the time series of functional brain networks and interpreted the resulted persistence landscape.

\section{Conclusion}
In this paper, some of the contributions on the application of persistence homology in time series and systems analysis have been surveyed. 

We started with the theoretical aspects of topological data analysis and introduced theoretical definitions and concepts (e.g., chain complexes, persistent homology, barcodes, persistent diagrams, persistent landscape, etc.).

For the application section, we reviewed the approaches that utilize the introduced theoretical concepts in real applications. We covered the topological analysis of dynamical systems where a great majority of works were in stability determination or periodic behaviour of signals and systems. It is not a surprise that the same techniques are applicable in risk analysis and predicting critical transitions in financial markets. This is where the topological data analysis shows its quality in detection early signs of crises from the market time series. 
The quality of TDA in this particular problem along with the importance of the problem, suggest that it can be a popular direction of research in the intersection of TDA and financial econometrics.

\bibliographystyle{unsrt}
\bibliography{main}

\end{document}